\documentclass[aps,prb,twocolumn,superscriptaddress,showpacs,floatfix]{revtex4}

\usepackage{graphicx}
\graphicspath{{figs/}}
\begin{document}
\title{Edge states induce boundary temperature jump in molecular dynamics simulation of heat conduction}
\author{Jin-Wu~Jiang}
    \affiliation{Department of Physics and Centre for Computational Science and Engineering,
             National University of Singapore, Singapore 117542, Republic of Singapore }
\author{Jie Chen}
    \affiliation{Department of Physics and Centre for Computational Science and Engineering,
                 National University of Singapore, Singapore 117542, Republic of Singapore }
\author{Jian-Sheng~Wang}
       \altaffiliation{Corresponding author: phywjs@nus.edu.sg}
    \affiliation{Department of Physics and Centre for Computational Science and Engineering,
                 National University of Singapore, Singapore 117542, Republic of Singapore }
\author{Baowen~Li}
        \affiliation{Department of Physics and Centre for Computational Science and Engineering,
                     National University of Singapore, Singapore 117542, Republic of Singapore }
        \affiliation{NUS Graduate School for Integrative Sciences and Engineering,
                     Singapore 117462, Republic of Singapore}
\date{\today}
\begin{abstract}
We point out that the origin of the commonly occurred boundary
temperature jump in the application of No\'se-Hoover heat bath in
molecular dynamics is related to the edge modes, which are
exponentially localized at the edge of the system. If heat baths are
applied to these edge regions, the injected thermal energy will be
localized thus leading to a boundary temperature jump. The jump can
be eliminated by shifting the location of heat baths away from edge
regions. Following this suggestion, a very good temperature profile
is obtained without increasing any simulation time, and the accuracy
of thermal conductivity calculated can be largely improved.
\end{abstract}

\maketitle

Computer simulations of thermal transport are normally done by
equilibrium and/or non-equilibrium molecular dynamics. In the latter
approach, the experimental condition is mimicked by setting up
temperature gradient across the system. The temperatures of two ends
are kept fixed. No\'se-Hoover heat bath is one of the most effective
approaches to realize constant temperature.\cite{Nose, Hoover} As a
generic phenomenon in existing works using No\'se-Hoover heat bath,
a significant boundary temperature jump (BTJ) occurs between the
temperature-controlled (TC) parts and the rest parts.\cite{Shiomi,
Wu, Zhang, WangSC} This temperature jump is usually regarded as the
consequence of thermal boundary resistance,\cite{Shiomi, WangSC} and
it results in both larger calculation error and longer simulation
time.

In this paper, we provide a sound physical explanation for the jump
by relating it to edge modes (EM), which will be fully excited but
localized on the boundary. Due to the EM, the thermal energy is
localized if heat bath is applied to these edge regions (which means
that the thermal current is injected into the system through these
edge regions), leading to the boundary temperature jump. We then
show that the temperature jump can be largely reduced by shifting
the location of heat baths away from edge regions. This shift
results in a better temperature profile, and gives rise to a more
accurate value of thermal conductivity. Our argument will be
illustrated by the case study of heat conduction in different
nanostructures such as nanoribon, nanotube and nanowire.

In our simulation, the second-generation Brenner inter-atomic
potential is used\cite{Brenner}. The Newton equations of motion are
integrated within the fourth order Runge-Kutta algorithm, in which a
time step of 0.5 fs is applied. The typical simulation time in
this paper is 8.5 ns.

\begin{figure}
  \begin{center}
    \scalebox{1.1}[1.1]{\includegraphics[width=7cm]{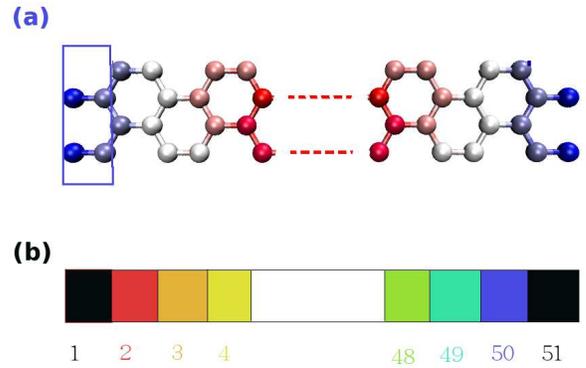}}
  \end{center}
  \caption{(Color online) (a). Configuration for the graphene nanoribbon with length 106~{\AA} and
  width 4.92~{\AA}. (b) is a schematic figure for (a). It is periodic in the vertical
  direction. The two out-most columns (column 1 and 51) are fixed.}
  \label{fig_cfg}
\end{figure}
Fig.~\ref{fig_cfg}~(a) is the configuration of a graphene nanoribbon in our
simulation. Each column contains four carbon atoms. There are 51 columns as shown
in the corresponding schematic diagram Fig.~\ref{fig_cfg}~(b). The out-most two black columns (1 and 51) are fixed. In vertical direction, periodic boundary condition is applied.

\begin{figure}
  \begin{center}
    \scalebox{1.1}[1.1]{\includegraphics[width=7cm]{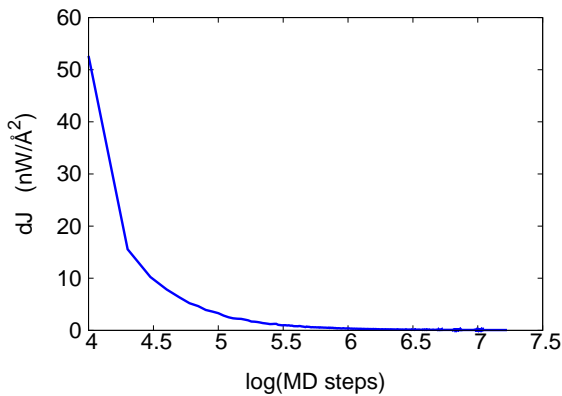}}
  \end{center}
  \caption{The difference between the heat current flows from the left
  and right heat baths. The heat current is averaged over 8.5 ns in the
  thermal steady state. The horizontal axis is plotted in log scale.}
  \label{fig_dJ}
\end{figure}
To study the thermal conductivity, we have to set up temperature
gradient across the system and then calculate the thermal conductivity
by Fourier's law. No\'se-Hoover heat baths are applied to
 columns 2 and 50 with temperatures 310 K and 290 K, respectively. $2\times 10^{7}$
 simulation steps (8.5 ns) are used for the system to reach thermal steady state.
 The difference between the thermal current from left and right heat baths, $dJ$,
 is used to determine whether the system has reached thermal steady state or not.
 In the steady state, $dJ$ should be zero. Fig.~\ref{fig_dJ} shows that the system reaches
 thermal steady state after $2\times 10^{7}$ steps where $dJ$ is almost zero.
 The nonzero value of $dJ$ is due to numerical errors and can be used to estimate the
  relative error of thermal conductivity as $dJ/J$, where $J$ is the heat current through the system.

\begin{figure}
  \begin{center}
    \scalebox{1.0}[1.0]{\includegraphics[width=7cm]{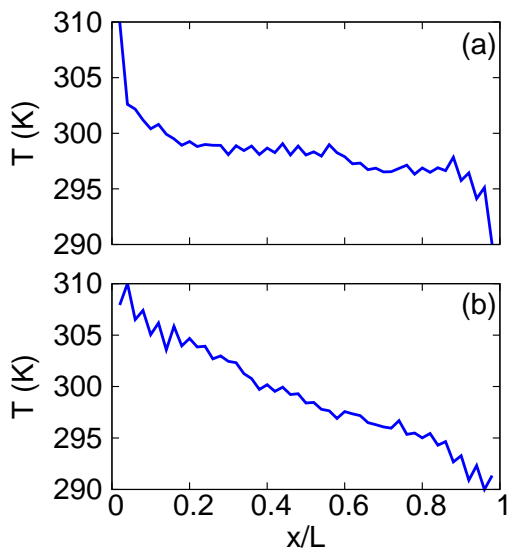}}
  \end{center}
  \caption{Temperature profiles at 300 K for graphene nanoribbon, whose configuration is shown in Fig.~\ref{fig_cfg}.
  (a). Heat baths are applied to columns 2 and 50 (edge regions).
  (b). Heat baths are applied to columns 3 and 49 (away from edge regions).
  The data in both figures is obtained by averaging over 8.5 ns in the steady state.}
  \label{fig_Tz}
\end{figure}

Fig.~\ref{fig_Tz}~(a) shows the temperature profile. The data is
obtained by averaging over 8.5 ns after the steady state is
achieved. This is a typical figure which also shows up in other
existing works.\cite{Shiomi, Zhang, WangSC} It shows that the
temperatures of the two TC parts are well controlled to be the
required value. But an obvious jump occurs between the TC parts and
the rest parts, i.e., between columns 2/3 on the left and columns
49/50 on the right. This jump can not be removed by simply
increasing simulation time and it leads directly to two negative
effects. Firstly, only temperatures in these columns far away from
edges can be used to do linear fitting to get temperature gradient.
The obtained value of temperature gradient is small ($-$5.5 K/\AA)
and sensitive to how many columns are chosen to do the linear
fitting. So error in the temperature gradient will be large.
Secondly, the thermal current $J$ is only 0.39 nW/\AA$^{2}$, which
is also very small. As a result, longer simulation time is needed
for the system to reach thermal steady state and the calculated
value for thermal conductivity (75.5 W/(mK)) has large relative
error estimated by $dJ/J=14.3\%$.

\begin{figure}
  \begin{center}
    \scalebox{1.1}[1.1]{\includegraphics[width=7cm]{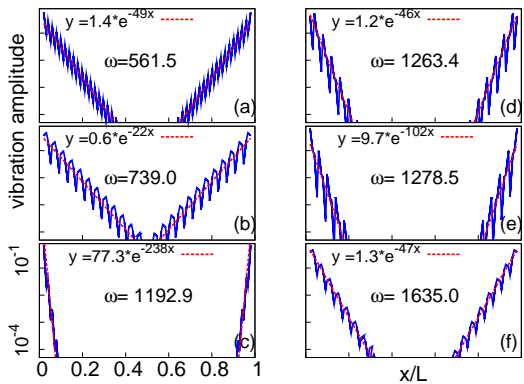}}
  \end{center}
  \caption{(Color online) Normalized vibration amplitudes vs.~reduced $x$ coordinate of each carbon atom. From (a) to (f) are six edge modes in graphene nanoribbon shown in Fig.~\ref{fig_cfg}. The frequency $\omega$ for each mode given in the figure is in cm$^{-1}$. The vertical axis is in log scale.}
  \label{fig_fixbc_edgestate}
\end{figure}

\begin{figure}
  \begin{center}
    \scalebox{1.0}[1.0]{\includegraphics[width=7cm]{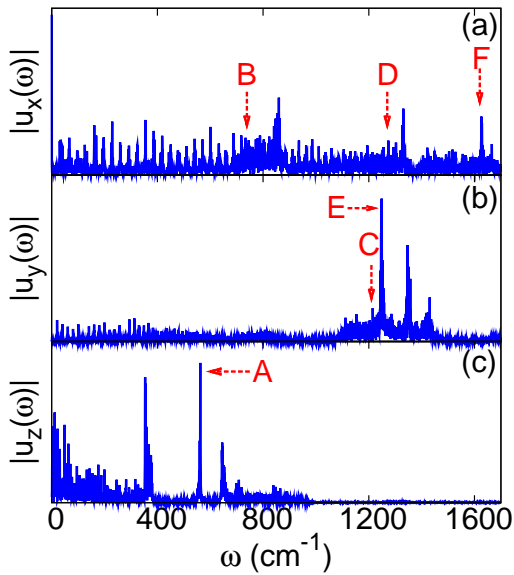}}
  \end{center}
  \caption{(Color online) Fourier transform $\vec{u}(\omega)$ of the vibrational amplitude $\vec{u}(t)$ of the atom in the edge regions (columns 2 and 50). Labels A, B, C, D, E, F denote the six corresponding localized edge modes in Fig.~\ref{fig_fixbc_edgestate}.}
  \label{fig_u_omega}
\end{figure}

To find an effective method to reduce this temperature jump, we
first have to understand the underlying mechanism. Geometrically,
the edge regions (columns 2/50) are very different from the other
regions inside the system. They are in the edge of the system, where
some eigen modes' vibrational amplitude decreases to zero very
quickly from edges into center. Fig.~\ref{fig_fixbc_edgestate} shows
the normalized vibrational amplitudes in all six EM in this system,
calculated from Brenner empirical potential implemented in `General
Utility Lattice Program'\cite{Gale}. It shows that the amplitudes
decrease exponentially (red dotted fitting line) and these edge
modes are doubly degenerate, since they can be localized at either
left or right edge regions (columns 2/50). Fig.~\ref{fig_u_omega}
shows explicitly that these EM have been excited. In
Fig.~\ref{fig_u_omega}, we do the Fourier transform for the
vibrational amplitude of one atom in column 2. The other atoms in
edge regions (columns 2 and 50) have similar result. Each peak
corresponds to an excited phonon mode. We have denoted the six fully
excited LEM in the figure. They are very important in two senses if
the thermal current is injected through these edge regions: (1). In
these modes, the vibration is localized at the edges, thus the
thermal current will be localized, leading to small thermal current
and small temperature gradient. (2). The total degrees of freedom in
the TC part is 12 (4 atoms) only, while there are six LEM. So the
LEM have a very large component, as half of degrees of freedom in
the TC parts are localized. Because of these two factors, large
amount of thermal current is localized at the left and right edge
regions. As a result, although the temperatures at the edges
(columns 2/50) can reach the required value, there will be a big
jump between TC parts and rest parts.

To reduce the temperature jump, one possible way is to increase the
number of atoms in the TC region. So that the component of the EM
will decrease relatively and the localized effect on the thermal
transport will be suppressed. As a result, the jump will be reduced.
The TC parts can be enlarged either in the vertical or longitudinal
direction. However, we find that the number of EM will also increase
if the TC part is enlarged in the vertical direction. Yet, this
number is kept to be a constant if the TC part is enlarged in the
longitudinal direction. So in this method, one can only enlarge the
TC part in the longitudinal direction to reduce BTJ. It has been
demonstrated that enlarging the system in the longitudinal direction
has some effects on reducing the jump.\cite{Shiomi}


Here we propose a more efficient method to solve this temperature
jump problem. Due to the localization property of the EM, they are
localized at the edge regions (columns 2/50), with the typical localization
length as $L_{loc}$=1 column. We can therefore shift the location of
heat baths to the regions where it is $L_{loc}$ away from the
boundary. Now we put columns 3 and 49 in the heat bath, instead of
columns 2 and 50. In this way, the thermal current can be
transported between TC parts (columns 3/49) and rest parts
efficiently. We mention that the six EM in the edge regions are
still excited in this situation. However, they can not generate
localization effect on the thermal current. Because now the thermal
current is injected through columns 3 and 49. So we can achieve a
very good temperature profile as shown in Fig.~\ref{fig_Tz}~(b). The
simulation time for this figure is the same as that in
Fig.~\ref{fig_Tz}~(a). Actually this temperature profile can be
obtained within much shorter simulation time (4.0 ns), since the
thermal current can now be injected into the system much more
efficiently. We get a much larger temperature gradient $-$16.6
K/\AA, and larger thermal current 0.96 nW/\AA$^{2}$. The obtained
thermal conductivity is 61.6 W/(mK) with relative error
$dJ/J=6.2\%$. This error is smaller than the previous one by a
factor of two. So following this proposal, one can calculate the
thermal conductivity much more accurately by changing the location
of heat baths away from the edge regions, which will not increase
any simulation time.

\begin{figure}
  \begin{center}
    \scalebox{0.9}[0.9]{\includegraphics[width=7cm]{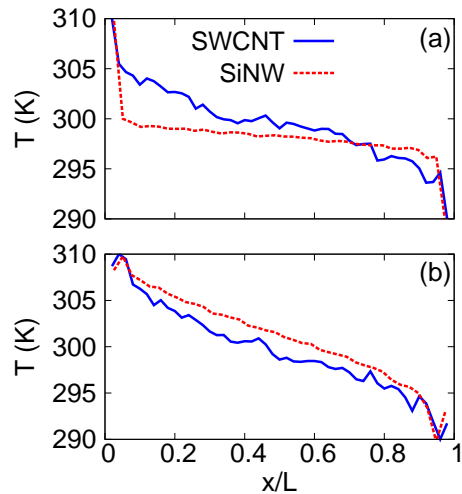}}
  \end{center}
  \caption{(Color online) Temperature profiles for SWCNT (blue solid) and SiNW (red dotted). (a). Heat
baths are applied to edge regions. (b). Heat baths are away from
edge regions.}
  \label{fig_Tz_SWCNT_SiNW}
\end{figure}
Our explanation for temperature jump is actually independent of
system. Because EM is essentially originating from the specific
geometrical configuration of edge regions.\cite{KimYS} This is
similar to electronic or spin edge states.\cite{Lima, YangL, Bermudez} To
check the generality of our method, we apply it to reduce BTJ in
single-walled carbon nanotube (SWCNT) and silicon nanowire (SiNW).
The results are shown in Fig.~\ref{fig_Tz_SWCNT_SiNW}. This figure
confirms the applicability of our method. For SWCNT (5, 0) with
length 106~{\AA} in this figure, we study the phonon modes in the
system and find 3 LEM with frequency as 761.5 (double degenerate),
1331.5 (double degenerate), 1416.0 (fourfold degenerate). Since
there are fewer LEM in SWCNT, BTJ in this system is not very large
even if heat baths are applied to edge regions. SiNW has a cross
section of $3\times 3$ unit cells (lattice constant 5.43~{\AA}) and
10 unit cells in the longitudinal direction. Compared with the
graphene nanoribbon or SWCNT, SiNW has a much larger BTJ if heat
baths are applied directly to the edge regions. This is because of
the high surface-to-volume ratio of SiNW. As a result, the number of
EM is very large. We find 30 EM in this SiNW, which is the reason
for large BTJ. This large BTJ can be reduced efficiently by changing
location of heat baths away from edge regions to avoid the
localization effect from EM (see Fig.~\ref{fig_Tz_SWCNT_SiNW}).
After the temperature jump is largely reduced, the temperature
gradient and thermal current across the system are enhanced by about
a factor of 7. So the thermal energy is efficiently pumped in, thus
reducing simulation time, and the calculation error will be smaller
by a factor of 7.

In conclusion, we have addressed a very generic problem - the
boundary temperature jump in molecular dynamics simulation of heat
conduction. We have provided strong evidence that this jump is
related to the EM, which are localized exponentially in the edge
regions. Theses modes will localize thermal current if heat baths
are applied exactly to these edge regions, leading to BTJ. An
effective way to reduce BTJ is to shift the location of heat baths
away from edge regions. In this way, the thermal current can be
transported between TC parts and rest parts efficiently and a better
temperature gradient can be established. Therefore, thermal
conductivity can be calculated much more accurately without
increasing computation time.

We have five further remarks: (1). Here we consider No\'se-Hoover
heat bath to illustrate that EM take the responsibility for
temperature jump. Actually, this explanation is also valid when
other heat bath is applied. Because the essential function of heat
bath is to excite all phonon modes in the TC part, including
localized modes. So in the case of other kind of heat bath,
temperature jump also occurs if heat bath is applied exactly to the
edge regions. Due to its randomness property at each step, the
Langevin heat bath can suppress some of the accumulation effect of
EM. So BTJ in Langevin is generally smaller than that in
No\'se-Hoover heat bath.\cite{ChenJ} (2). In this paper, we use fixed boundary
condition in the longitudinal direction. For free boundary
condition, we also find eight doubly degenerate EM with frequency as
97.3, 244.4, 473.4, 516.5, 1379.3, 1634.4, 1697.7, and 1703.0
cm$^{-1}$. These modes are localized at columns 1 and 51, which are
edge regions in free boundary condition. In this case, the number of
EM is increased and the frequencies for the first four are lower. So
the localization effect will be more serious if heat baths are
applied to the edge regions (columns 1 and 51), resulting in even
larger BTJ.\cite{Zhang} BTJ in this situation can also be removed by
changing the location of the heat baths away from the edge regions.
For periodic boundary condition, the EM turn into optical phonon
modes in the system. The velocity of optical phonon modes is very
small, which will also have `localization' effect on the thermal
current, leading to big BTJ.\cite{Schelling} 
This BTJ can be reduced by artificially confining the injected thermal
energy to acoustic phonon modes.\cite{Keblinski} However, practically it is more
complicated, since the injected phonon should follow the Bose-Einstein
statistics if the thermal conductivity is studied.
(3). In experiments, these EM will also be
excited if the heat source or sink is located exactly in the edge
regions, leading to smaller thermal current and smaller temperature
gradient. As a result, experimental errors increase. So it is
important to put heat source and sink away from the edge regions to
avoid the thermal current being localized. (4). BTJ
can be removed by enlarging TC parts in the longitudinal direction
as in Ref.~\onlinecite{Shiomi}, or by changing location of heat bath
as in this paper. Actually there is a small remaining BTJ in both
results (Fig. 2 in Ref.~\onlinecite{Shiomi} and
Fig.~\ref{fig_Tz}~(b), Fig.~\ref{fig_Tz_SWCNT_SiNW}~(b) in this
paper). This small BTJ is the result of thermal boundary resistance
between TC parts and rest parts. How to remove this small remaining
BTJ is still a problem and requires further investigation.
(5). The EM discussed in this paper has some similarity to
the Rayleigh waves,\cite{Rayleigh} which is confined to travel across surfaces of solids.
The wave velocities of Rayleigh waves are slow, leading to some `localization'
effect. Yet, the penetrate depth of Rayleigh is approximately equal
to the wavelength, which is larger than that of the EM (typically $L_{loc}\approx1$ column, 
in Fig.~\ref{fig_fixbc_edgestate}).

\textbf{Acknowledgements}. We thank Prof. Pawel Keblinski for
helpful comments, and Xiang Wu and Nuo Yang for useful discussions.
The work is supported by a Faculty Research Grant No.
R-144-000-173-101/112 of NUS, and Grant No. R-144-000-203-112 from
Ministry of Education of Republic of Singapore, and Grant No.
R-144-000-222-646 from NUS.


\begin{thebibliography}{}
\bibitem{Nose} S. No\'se, J. Chem. Phys. \textbf{81}, 511 (1984).

\bibitem{Hoover} W. G. Hoover, Phys. Rev. A, \textbf{31}, 1695 (1985).

\bibitem{Shiomi} J. Shiomi and S. Maruyama, Japan. Soc. Appl. Phys. \textbf{47}, 2005 (2008).

\bibitem{Wu} M. C. H. Wu and J. Y. Hsu, Nanotechnology \textbf{20}, 145401 (2009).

\bibitem{Zhang} G. Zhang and B. Li, J. Chem. Phys. \textbf{123}, 114714 (2005).

\bibitem{WangSC} S.-C. Wang, X.-G. Liang, X.-H. Xu, and T. Ohara, J. Appl. Phys. \textbf{105}, 014316 (2009).

\bibitem{Brenner} D. W. Brenner, O. A. Shenderova, J. A. Harrison, S. J. Stuart, B. Ni and S. B. Sinnott, J. Phys.:Condens. Matter \textbf{14}, 783 (2002).

\bibitem{Gale} J. D. Gale, JCS Faraday Trans., \textbf{93}, 629 (1997).

\bibitem{KimYS} S. Y. Kim and H. S. Park, Nano. Lett. \textbf{9}, 969 (2009).

\bibitem{Lima} M. P. Lima, A. Fazzio, and A. J. R. da Silva1, Phys. Rev. B, \textbf{79}, 153401 (2009).

\bibitem{YangL} L. Yang, M. L. Cohen, and S. G. Louie, Phys. Rev. Lett. \textbf{101}, 186401 (2008).

\bibitem{Bermudez} A. Bermudez, D. Patane, L. Amico, and M. A. M. Delgado, Phys. Rev. Lett. \textbf{102}, 135702 (2009).
                   
\bibitem{ChenJ} J. Chen, G. Zhang, and B. Li (unpublished).

\bibitem{Schelling} P. K. Schelling, S. R. Phillpot, and P. Keblinski, Phys. Rev. B, \textbf{65}, 144306 (2002).

\bibitem{Keblinski} Private communication with Pawel Keblinski.

\bibitem{Rayleigh} L. Rayleigh, Proc. London Math. Soc. \textbf{17}, 4 (1885).


\end{thebibliography}
\end{document}